# A Novel Sliding Mode Control for a Class of Affine Dynamic Systems

Zu-Ren Feng, *Member, IEEE*, Rui-Zhi Sha, Na Lu, and Chen-Long He

*Abstract*—This paper proposes a novel sliding mode control (SMC) method for a class of affine dynamic systems. In this type of systems, the high-frequency gain matrix (HFGM), which is the matrix multiplying the control vector in the dynamic equation of the sliding variables vector, is neither deterministic nor positive definite. This case has rarely been covered by general SMC methods, which perform well under the condition that the HFGM is certain or uncertain but positive definite. In this study, the control law is determined by solving a nonlinear vector equation instead of the conventional algebraic expression, which is not applicable when the HFGM is uncertain and non-positive definite. Theorems with some relaxed system parametric uncertainty assumptions are proposed to guarantee the existence and uniqueness of the solution, and proofs of them, based on the principle of the convex cone set, are given in the text. The proposed control strategy can be easily applied in practice, and the chattering caused by the discontinuous control can be suppressed, as it can in general SMCs. The proposed controller was used in two affine dynamic systems, and the simulation results demonstrate its effectiveness.

## I. Introduction

AN affine dynamic system usually refers to a system whose dynamics is affine with respect to control [1], [3], [10], [11], [18]. It covers many mechanical systems [24], [25], [26], [32], such as power systems [33]; robot dynamics [27], [28]; helicopter systems [29], [30]; and chemical control systems [31], etc.. Many strategies are raised to control this class of systems [25], [34], [35], [36], [37], [38], one of which is the sliding mode control (SMC) [5], [6], [7], [8], [9], [11], [12], [13], [14], [15], [16], [18], [20], [21], [22],[23].

The SMC concept is derived from the variable structure system theory that appeared in the middle of the twentieth century. In variable structure systems, the control law is allowed to change its structure [1], and this characteristic makes it possible to combine useful properties of each structure and then generate new properties not existing in any of the original structures. This idea led to a variable structure control method, the SMC, which was created and developed after the 1950s and popularized by Utkin [2]. In this method, an exponentially stable sliding surface [10], [11] is defined as a function of tracking errors[12], and the Lyapunov theory ensures that all the states reach the surface, then the system will be asymptotically stable. An SMC has several advantages, such as robustness against disturbance or parametric uncertainties and no need for detailedsystems information.

When designing the control law of SMC, the first step is to define a sliding variables vector. In affine dynamic systems, the first-order derivative of the sliding variables vector is driven by the system input through the high-frequency gain matrix (HFGM). Pisano and Usai [3] divided affine dynamic systems into two classes: 1) where the gain matrix is known, and 2) where the gain matrix is uncertain but positive definite, and the control laws are put forward accordingly. Those two kinds of conditions have been the main concern of many researchers and have been well solved by them. Utkin [19] mentioned the uncertainty of the system matrix caused by variations in the parameters of a linear system; he demonstrated that an SMC has the ability to be insensitive to parameter variations. Yoo and Ham [20] proposed an adaptive SMC scheme where fuzzy logic systems are used to approximate the unknown HFGM to a sum of two positive matrices, one known and the other uncertain. As the research developed, many researchers began to study systems whose gain matrices were unknown and nonpositive and devised some effective control strategies. Nasiri et al. [21] proposed an adaptive SMC that did not require prior knowledge of the uncertainty bounds for multi-input/multioutput nonlinear systems, but the constraints of the model parameters are too strong for many systems to meet. Mobayen [22] proposed an adaptive chattering-free PID SMC law that can make the system states reach the sliding manifold in finite time under a condition of parametric uncertainties. However, the conclusion is wrong if there is no constraint on the update gains; a situation referred to in [39], [40].

In this paper, a novel SMC strategy with some relaxed system parametric uncertainty assumptions is proposed to solve the problem where the HFGM is uncertain and nonpositive definite. The control law is determined by solving a nonlinear vector equation, and the existence and uniqueness of the solution are proved based on the principle of the convex cone set. The algorithmic process for computing the control law is provided and is easy to implement in practice. The chattering caused by discontinuous control can be suppressed, as it can in general SMCs, and two applications demonstrate the effectiveness of the proposed method.

The paper is organized as follows: after some notations are listed, in Section 2 a general form of a class of SMC systems is presented. Then in Section 3, the new method is proposed, including the nonlinear vector equation from which the control

Z.-R. Feng, R.-Z. Sha, N. Lu and C.-L. He are with the State Key Laboratory for Manufacturing Systems Engineering, Systems Engineering Institute, Xi'an Jiaotong University, Xi'an 710049, China (e-mail: fzr9910@xjtu.edu.cn; ruizhi.sha@gmail.com;lvna2009@xjtu.edu.cn;chenlong.he@stu.xjtu.edu.cn).

law is to be solved, the lemmas and theorems by which the existence and uniqueness of the solution are given theoretically, and the calculation process of the control law. In Section 4, two multi-input applications are presented, and finally some conclusions are drawn. Detailed proofs of the lemmas in Section 3 are included in the Appendices.

*Notation.* We denote by $I_n$ the $n$-dimensional identity matrix. The set $\{1,...,n\}$ is denoted by $[n]$. Given a vector $x \in \mathbb{R}^n$, we denote by $[x]_i$ the $i$-th element of $x$, and denote by $|x| \in \mathbb{R}^n$ the absolute vector of $x$ with $[|x|]_i = |[x]_i|$, $i \in [n]$. Given a matrix $D \in \mathbb{R}^{n \times m}$, we denote by $[D]_{ij}$ the element in the $i$-th row and $j$-th column of $D$. We denote by $\Sigma^n$ the $n \times n$ diagonal matrix set $\Sigma^n = \{S \mid S = diag\{\zeta_1,...,\zeta_n\}, \zeta_i \in \{-1,1\}\}$. We define the matrix function $S(x): \mathbb{R}^n \to \Sigma^n$ with its diagonal elements being the symbolic function of an $n$-dimensional vector $x$; that is, $S(x) = diag\{sign([x]_i)\}$, where
$$sign([x]_i) = \begin{cases} 1, & [x]_i \geq 0 \\ -1, & [x]_i < 0 \end{cases}.$$

## II. PREREQUISITE KNOWLEDGE

We discuss a class of affine dynamic systems that can be expressed in the form
$$\dot{x}(t) = A(x(t),t) + B(x(t),t)u(t) + \xi(t), \quad (1)$$
where $x \in \Omega_x \subset \mathbb{R}^n$ is the state vector, $\Omega_x$ is an area containing the origin, $A: \mathbb{R}^n \times \mathbb{R}^+ \to \mathbb{R}^n$ is a vector field in the state space, $B: \mathbb{R}^n \times \mathbb{R}^+ \to \mathbb{R}^{n \times m}$ is the system control gain matrix, $u \in \mathbb{R}^m$ is the bounded control input vector, and $\xi \in \mathbb{R}^n$ is the bounded disturbance vector.

Define a set of sliding surfaces in the state space passing through the origin to represent a sliding manifold as follows:
$$\sigma(x) = 0, \quad (2)$$
where $\sigma: \mathbb{R}^n \to \mathbb{R}^m$ is a sufficiently smooth vector function and is designed so that the state $x$ is asymptotically stable if $\sigma(x) \to 0$.

Consider the $m$-dimensional vector
$$s = \sigma(x), \quad (3)$$
usually named a sliding variables vector [3]. The derivative of it is
$$\begin{aligned}\dot{s} &= J_x^\sigma(x)\dot{x}(t) \\ &= J_x^\sigma(x)A(x,t) + J_x^\sigma(x)B(x,t)u(t) + J_x^\sigma(x)\xi(t)\end{aligned}, \quad (4)$$
where $J_x^\sigma(x) = \partial\sigma(x)/\partial x^T$. For simplicity, hereafter the bracket behind $x$ in $A(x(t),t)$ and $B(x(t),t)$ is omitted.

Let $f(x,t) = J_x^\sigma(x)A(x,t)$, $G(x,t) = J_x^\sigma(x)B(x,t)$, and $\eta(t) = J_x^\sigma(x)\xi(t)$. Then (4) can be written as
$$\dot{s} = f(x,t) + G(x,t)u(t) + \eta(t), \quad (5)$$
where the vector $f(x,t)$ and the HFGM $G(x,t)$ are uncertain, and $\eta(t)$ is a bounded disturbance vector.

Equation (5) represents the general form of a class of sliding mode dynamic systems [3]. If the uncertain HFGM $G(x,t)$ is a symmetric positive definite and the lower bound of its eigenvalues is known, it is easy to obtain a control law that makes $s$ asymptotically stable [3]. Obviously, this kind of restriction on $G(x,t)$'s uncertainty is too strict. In the relevant research, the descriptive form in the literature of robust control [41] is used for reference; that is, $G(x,t) = G_0(x,t) + \Delta G(x,t)$, where $G_0(x,t)$ is the known part and $\Delta G(x,t)$ the uncertain part. In [21], the uncertain HFGM $G(x,t)$ is required to satisfy the uncertainty constraints $\left|[F(x,t)]_{ij}\right| < 1/m, \forall i,j \in [m]$, where $F(x,t) = \Delta G(x,t)G_0^{-1}(x,t)$. In Section 3, we propose a novel SMC strategy with relaxed constraints on the uncertainty of $G(x,t)$.

## III. THE NOVEL SLIDING MODE CONTROL METHOD

### A. Basic theories

Assume that (5) can be expressed as
$$\begin{aligned}\dot{s} &= (f_0(x,t) + \Delta f(x,t)) \\ &\quad + (G_0(x,t) + \Delta G(x,t))u(t) + \eta(t)\end{aligned}, \quad (6)$$
where $f_0(x,t)$ and $G_0(x,t)$ represent nominal terms of $f(x,t)$ and $G(x,t)$, and $\Delta f(x,t)$ and $\Delta G(x,t)$ represent the uncertain terms of $f(x,t)$ and $G(x,t)$ respectively.

**Assumption 1.** *For all $x \in \Omega_x$ and $t \geq 0$, the matrix $G_0(x,t)$ is nonsingular.*

**Assumption 2.** *For all $x \in \Omega_x$ and $t \geq 0$, the uncertain vectors $\Delta f(x,t)$ and $\eta(t)$ satisfy the following boundedness conditions:*
$$\begin{cases} |\Delta f(x,t)| \leq \bar{f}(x,t) \\ |\eta(t)| \leq \bar{\eta} \end{cases}, \quad (7)$$
*where $\bar{f}(x,t)$ and $\bar{\eta}$ are nonnegative known vectors.*

The effect of $\Delta G(x,t)$ on the system depends on its proportion in the whole gain matrix, so the uncertainty degree $\Delta G(x,t)$ should be characterized in relative to $G_0(x,t)$. In this paper, a general description form is adopted. Let $G_0(x,t)$ be decomposed into a product of two $m \times m$ nonsingular matrices in the form
$$G_0(x,t) = M(x,t)Q(x,t). \quad (8)$$
Let
$$F(x,t) = M^{-1}(x,t)\Delta G(x,t)Q^{-1}(x,t), \quad (9)$$
where $F(x,t)$ is an uncertain matrix, and $M(x,t)$ and $Q(x,t)$ are known matrices that reflect the structural information of the uncertain part of $G(x,t)$. Then the HFGM can be expressed as
$$G(x,t) = M(x,t)(I_m + F(x,t))Q(x,t). \quad (10)$$

**Assumption 3.** For all $x \in \Omega_x$ and $t \geq 0$, the uncertain matrix $F(x,t)$ satisfies the boundedness condition

$$\left|[F(x,t)]_{ij}\right| \leq [\overline{F}]_{ij}, \forall i,j \in [m], \quad (11)$$

where $\overline{F}$ is a nonnegative matrix called the upper bound matrix (UBM) of $F(x,t)$.

Substituting (10) into (6), the dynamic equation of $s$ can be written as

$$\dot{s} = f_0(x,t) + \Delta f(x,t) + M(x,t)(I_m + F(x,t))\hat{u}(t) + \eta(t) \quad (12)$$

where

$$\hat{u}(t) = Q(x,t)u(t). \quad (13)$$

Let $v = M(x,t)s$ and $S(s)$ and $S(v)$ be $m \times m$ matrix functions as defined in the Notation.

**Theorem 1.** *The sliding variable vector $s$ in (12) is asymptotically stable if the control $\hat{u}(t)$ satisfies the vector equation*

$$\hat{u}(t) = -M^{-1}(x,t)\left(f_0(x,t) + S(s)(\overline{f}(x,t) + \overline{\eta}) + \rho s\right)$$
$$-S(v)\overline{F}(x,t)|\hat{u}(t)| \quad . \quad (14)$$

**Proof.** Define a positive definite function

$$V(s) = 1/2 s^T s. \quad (15)$$

Along the trajectory of (12), the derivative of $V(s)$ with respect to time is given by

$$\dot{V}(s) = s^T (f_0(x,t) + \Delta f(x,t))$$
$$+ s^T (M(x,t)(I + F(x,t))\hat{u}(t) + \eta(t)) \quad . \quad (16)$$

Considering Assumption 2, (16) can be deduced to the following inequality:

$$\dot{V}(s) \leq s^T f_0(x,t) + |s|^T (\overline{f}(x,t) + \overline{\eta})$$
$$+ s^T M(x,t)(F(x,t)\hat{u}(t) + \hat{u}(t)) \quad . \quad (17)$$

Substituting (14) into (16), and considering Assumption 3, we get

$$\dot{V}(s) \leq s^T f_0(x,t) + |s|^T (\overline{f}(x,t) + \overline{\eta})$$
$$- s^T (f_0(x,t) + S(s)(\overline{f}(x,t) + \overline{\eta}) + \rho s)$$
$$+ v^T (F(x,t)\hat{u}(t) - S(v)\overline{G}(x,t)|\hat{u}(t)|) \quad . \quad (18)$$
$$\leq -\rho s^T s + v^T F(x,t)\hat{u}(t) - |v|^T \overline{G}(x,t)|\hat{u}(t)|$$
$$\leq -\rho s^T s < 0$$

Therefore, the origin of the $m$-dimensional space of variables $s$ is an asymptotically stable equilibrium point. □

Because the control $\hat{u}(t)$ is solved from the vector equation, which is nonlinear with respect to $\hat{u}(t)$, the key problem is, From (14) is it possible to get a unique solution?

If $\overline{F}$ were a zero matrix, (14) would become a formula from which $\hat{u}(t)$ could be calculated directly. However, this means that $\Delta G(x,t) = 0$; that is, $G(x,t)$ is deterministic.

Otherwise, let

$$u_c = -M^{-1}(x,t)\left(f_0(x,t) + S(s)(\overline{f} + \overline{\eta}) + \rho s\right). \quad (19)$$

Then (14) can be expressed as

$$\hat{u}(t) + S(v)\overline{F}|\hat{u}(t)| = u_c. \quad (20)$$

To solve the problem conveniently, it is better to change (20) into a compact form as follows (in the statements of lemmas and theorems below, the arguments $x$ and $t$ in brackets are omitted for simplicity):

$$(S + H)p = u_c, \quad (21)$$

where $H = S(v)\overline{F}$, $S = S(\hat{u})$, and $p = |\hat{u}| > 0$. Therefore, solving $\hat{u}$ from (20) is equivalent to solving a pair of $p \geq 0$ and $S \in \Sigma^m$ from (21), and the solution of (20) is calculated by the formula $\hat{u} = Sp$.

According to the definition of $S$, there is a total of $2^m$ different selections of $S$, named $S_i, i \in [2^m]$, and let $D_i = S_i + H$.

**Definition 1.** *Define $2^m$ closed convex cones in the $m$-dimensional space $\mathbb{R}^m$ in the form*

$$C_i = \{D_i p | \forall p \geq 0\}, i \in [2^m]. \quad (22)$$

*Correlatively, the interior open region and the conical surface of $C_i$ are denoted by $C_i^o = \{D_i p | \forall p > 0\}$ and $\widehat{C}_i = C_i - C_i^o$ respectively.*

About the convex cones defined by Definition 1, we have the following three lemmas:

**Lemma 1.** *If $\|H\| < 1$, then $C_i^o \cap C_j^o = \varnothing$ for all $i \neq j$.*

**Proof.** See Appendix A.

**Lemma 2.** *Any closed convex cone $C_i, i \in [2^m]$ does not have an independent conical surface; that is, if an $m$-dimensional vector $y \in \widehat{C}_i$, then there exists $\widehat{C}_j, j \neq i$ such that $y \in \widehat{C}_j$.*

**Proof.** See Appendix B.

**Lemma 3.** *If $\|H\| < 1$, then $C = \bigcup_{i=1}^{2^m} C_i$ is equal to $\mathbb{R}^m$.*

**Proof.** See Appendix C.

Let $\|\overline{F}\|$ be any induced matrix norm of $\overline{F}$. The theorem below gives the condition of $\|\overline{F}\|$ to ensure the existence as well as the uniqueness of solution $\hat{u}$ for (20).

**Theorem 2.** *Consider (20) with matrix $S(v) \in \Sigma^m$ and vector $u_c \in \mathbb{R}^m$ being arbitrarily given. The solution $\hat{u} \in \mathbb{R}^m$ exists and is uniqu if $\|\overline{F}\| < 1$.*

**Proof.** If $u_c = 0$, because $\|H\| = \|S(v)\|\|\overline{F}\| = \|\overline{F}\| < 1$, then $S + H$ in (21) is nonsingular, so there exists a unique solution $p = 0$; therefore, $\hat{u} = Sp = 0$ is unique.
If $u_c \neq 0$, the proof will be divided into two parts.

**Proof of existence**
According to Lemma 3, any $u_c \in \mathbb{R}^m$ belongs to at least one closed convex cone, say $C_i$, and then it can be expressed as $u_c = (S_i + H)p$ with $p \geq 0$, hence the solution $\hat{u} = S_i p$ exists.

**Proof of uniqueness**

- If $u_c \in C_i^o$, according to Lemma 1, $u_c \notin C_j^o$, $\forall j \neq i$, then $u_c$ is expressed uniquely by $u_c = (S_i + H)p$ with $p > 0$, and $\hat{u} = S_i p$ is unique.
- If $u_c \in \hat{C}_i$, then $u_c = (S_i + H)p$ has the unique solution $p \geq 0$, and at least one element of $p$ is zero. According to the proof of Lemma 2 (See Appendix B), $u_c$ is also on the conical surface $\hat{C}_j$ as long as $S_j p = S_i p$. It follows that the solution $\hat{u} = S_i p = S_j p$ is naturally unique.

□

If $\bar{F}$ is a symmetric matrix, according to the properties of the spectral radius of a matrix [42], its 2-norm is equal to its spectral radius, and is therefore the lower bound of all the induced norms of $\bar{F}$; that is, $\|\bar{F}\|_2 \leq \|\bar{F}\|$. In this case, it is more appropriate to use the 2-norm of $\bar{F}$ in Theorem 2. Moreover, the next theorem shows that, if $\bar{F}$ is symmetric, the condition $\|\bar{F}\|_2 < 1$ is also necessary for (20) to have a unique solution.

**Theorem 3.** For (20) with matrix $S(v) \in \Sigma^m$ and vector $u_c \in \mathbb{R}^m$ being arbitrarily given, if matrix $\bar{F}$ is symmetric, the solution $\hat{u} \in \mathbb{R}^m$ exists and is unique, if and only if $\|\bar{F}\|_2 < 1$.

**Proof.** The proof of sufficient condition is the same as that of Theorem 2; here, the disproof is adopted to demonstrate the necessity of the condition $\|\bar{F}\|_2 < 1$.

Because $\bar{F}$ is a symmetric matrix, all eigenvalues of $\bar{F}$ are real, and their absolute values are bounded by $\|\bar{F}\|_2$. For any eigenvalue $\lambda$ of $\bar{F}$, it will be proved that if $|\lambda| \geq 1$, there exists a pair of $S(v)$ and $u_c$, which makes the solution of (20) not unique.

Notice that $S(v)\bar{F}S(v)$ and $\bar{F}$ have the same eigenvalues. Let $y$ be the eigenvector of $S(v)\bar{F}S(v)$ corresponding to $\lambda$; that is,

$$S(v)\bar{F}S(v)y = \lambda y . \quad (23)$$

Assuming that $S(v) = S(y)$, then the nonzero vector $S(v)y$, denoted by $p$, is also nonnegative. Adding $Sp$, $S \in \Sigma^m$ to both sides of (23), we get

$$(S + S(v)\bar{F})p = (S + \lambda S(v))p . \quad (24)$$

Because $p$ is a nonnegative vector, assuming that $\hat{u} = Sp$ and $u_c = (S + \lambda S(v))p$, it is easy to see that (24) and (20) are equivalent.

If $|\lambda| = 1$, let $S = -\lambda S(v)$, and then (24) becomes $(S + S(v)\bar{F})p = 0$. It follows that for $u_c = 0$, (24) has two solutions, $\hat{u} = 0$ and $\hat{u} = -\lambda S(v)p \neq 0$.

If $|\lambda| > 1$, let $S(\sigma) = \sigma \, sign(\lambda) S(v), \sigma = -1$ or $1$ and substitute $S$ by $S(\sigma)$ in (24). We get

$$(S(\sigma) + S(v)\bar{F})p(\sigma) = u_c , \quad (25)$$

where $p(\sigma) = \dfrac{\lambda}{\sigma \, sign(\lambda) + \lambda} p$ and $u_c = \lambda S(v)p$.

Because $|\lambda| > 1$, it is easy to verify that whether $\sigma$ equals $1$ or $-1$, the vector $p(\sigma)$ is always nonnegative. Thus for a single $u_c = \lambda S(v)p$, (24) has two different solutions,
$\hat{u} = |\lambda| S(v)p/(\lambda + sign(\lambda))$ and
$\hat{u} = -|\lambda| S(v)p/(\lambda + sign(\lambda))$.

□

**Remark 1.** The reason for decomposing $G_0(x,t)$ in the form of (8) is that to avoid elements of $\hat{u}$ in (14) growing too large, it is necessary to search for a suitable pair of $M$ and $Q$ to make $\|\bar{F}\|$ as small as possible. If $G_0(x,t)$ is with non-negligible alteration caused by changes of $x$ and $t$, the selection of $G_0(x,t)$'s decomposition can be simply set as $M(x,t) = G_0(x,t)$, $Q(x,t) = I_m$ or $M(x,t) = I_m$, $Q(x,t) = G_0(x,t)$, depending on which $\|\bar{F}\|$ is smaller for all possible $G_0(x,t)$ and $\Delta G(x,t)$. Otherwise, $G_0$ can be chosen as a constant matrix, and through applying certain optimization algorithms (such as *particle swarm optimization (PSO)*), a pair of matrices $M$ and $Q$, which minimizes the norm of $\bar{F}$ for all possible $\Delta G(x,t)$, can be obtained.

**Remark 2.** When the decomposition of (8) is chosen as $M(x,t) = I_m$, $Q(x,t) = G_0(x,t)$, $F(x,t)$ is consistent with that in [21], then the condition in [21] is equivalent to $\max_{i,j}[\bar{F}]_{ij} < 1/m$. It is easy to prove that the condition $\|\bar{F}\| < 1$ can be derived from the condition $\max_{i,j}[\bar{F}]_{ij} < 1/m$, but the reverse is not necessarily true if $m > 1$; for example, $\bar{F} = 0.9 I_m$. So the condition in this paper is much weaker than that in [21].

*B. Calculation of the control law*

As presented before, the proposed SMC is obtained from solving a nonlinear vector equation. In practical applications, the algorithm should meet the requirements of real-time computing. Generally, the dimension $m$ of control is not very high, so (20) can be solved through enumerating $S$ by $2^m$ times at most; the time cost of the algorithm is not high. The calculation process of the control law is as follows:

Step 1：Calculate $u_c$ according to (19), $v = M(x,t)s$ and $H = S(v)\overline{F}$.

Step 2：Enumerate $S \in \Sigma^m$.

Step 3：Calculate $p = (S+H)^{-1} u_c$.

Step 4：Check the condition $S = \text{diag}\{\text{sign}([p]_i)\}$ ? if Yes, go to Step 5；if No, return to Step 2.

Step 5：Calculate $u = Q^{-1}(x,t)Sp$.

The system chattering caused by the discontinuous control from (14) can be suppressed, as it can in general SMCs. Rewrite (14) as

$$\hat{u}(t) - S(v)\overline{F}|\hat{u}(t)| = \\ -M^{-1}(x,t)\left(f_0(x,t) + S(s)(\overline{f}(x,t)+\overline{\eta}) + \rho s\right). \quad (26)$$

Assume that all elements of $M(x,t)$, $Q(x,t)$, $f_0(x,t)$, and $\overline{f}(x,t)$ are continuous functions of arguments. Because the control input is assumed to be bounded, the states of system (1) are continuous time functions. Simply replace the sign functions in $S(v)$ and $S(s)$ with continuous functions, it is easy to verify that the system control law $u(t) = Q^{-1}(x,t)\hat{u}(t)$ is continuous, so the chattering phenomenon will be suppressed.

Replace $S(x) = \text{diag}\{\text{sign}([x]_i)\}$ ($x$ refers to $s$ or $v$) with typical smoothing functions as below:

$$\tilde{S}_x = \text{diag}\{\mu(x_i, \delta_x)\},\\ \mu(x_i, \delta_x) = \begin{cases} \text{sign}(x_i), |x_i| \geq \delta_x \\ x_i/\delta_x, |x_i| < \delta_x \end{cases}, \quad (27)$$

where $\delta_x$ is the positive number relevant to $x$. When choosing $\delta_x$, reducing control errors and suppressing chattering should be considered in a compromised way.

## IV. EXPERIMENTS

### A. A liquid-filled spacecraft

*1) Model analysis and controller design*

A spacecraft filled with liquid fuel is a multi-input nonlinear system and is shown in Fig. 1. More details of its configuration can be found in [4] and [43].

The simplified dynamic model of the spacecraft is expressed as

$$\begin{bmatrix}\dot{v}_z \\ \ddot{\theta} \\ \ddot{\psi}\end{bmatrix} = N^{-1}G_x + N^{-1}G_u u, \quad (28)$$

where $\theta$ is the attitude angle, $\psi$ the fuel sloshing angle, $v_z$ the transverse velocity, $u = [f\ M]^T$ the input vectors, $f$ the transverse reaction jet force, and $M$ the pitching moment. The parameter matrices $N$, $G_x$, and $G_u$ are

$$G_x = \begin{bmatrix} (m+m_f)\dot{\theta}v_x + m_f a(\dot{\theta}+\dot{\psi})^2 \sin\psi \\ mb\dot{\theta}v_x \\ -\varepsilon\dot{\psi} - \dfrac{m_f aF}{m+m_f}\sin\psi + m_f a\dot{\theta}v_x \cos\psi \end{bmatrix},$$

$$G_u = \begin{bmatrix} 1 & 0 \\ b & 1 \\ 0 & 0 \end{bmatrix}, \text{ and}$$

$$N = \begin{bmatrix} m+m_f & m_f a\cos\psi + mb & m_f a\cos\psi \\ mb & I+mb^2 & 0 \\ m_f a\cos\psi & I_f + m_f a^2 & I_f + m_f a^2 \end{bmatrix},$$

where $m$ is the mass of the spacecraft, $I$ the moment of inertia (without fuel), $m_f$ and $I_f$ the fuel mass and moment of inertia respectively, $a$ the length of the pendulum, $b$ the distance between the pendulum point of attachment and the spacecraft center of mass location along the $x$-axis, and $F$ the thrust.

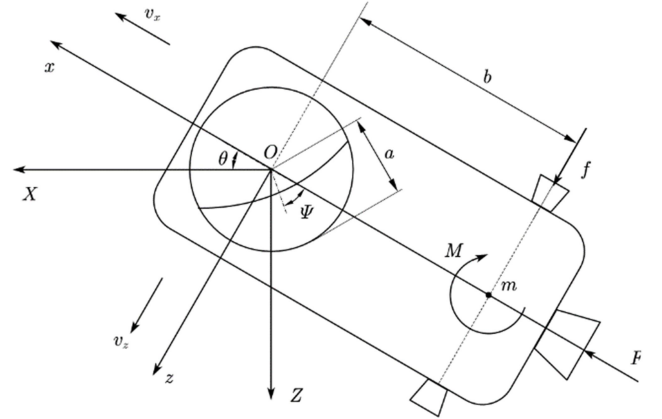

Fig. 1. Spacecraft with fuel slosh model

The control target is to guarantee the stability of the spacecraft and attenuate the fuel slosh; in other words, make $\theta$, $\psi$, and $v_z$ tend to zero rapidly.

Letting the state vector be

$$x = [v_z\ \theta\ \dot{\theta}\ \psi\ \dot{\psi}]^T, \quad (29)$$

the system is rewritten in state-space form as

$$\dot{x} = A(x) + B(x)u, \quad (30)$$

where

$$[A(x)]_2 = [x]_3, [A(x)]_4 = [x]_5,$$
$$[B(x)]_{21} = [B(x)]_{22} = [B(x)]_{41} = [B(x)]_{42} = 0,$$

$$\begin{bmatrix}[A(x)]_1 \\ [A(x)]_3 \\ [A(x)]_5\end{bmatrix} = N^{-1}G_x, \text{ and}$$

$$\begin{bmatrix}[B(x)]_{11} & [B(x)]_{12} \\ [B(x)]_{31} & [B(x)]_{32} \\ [B(x)]_{51} & [B(x)]_{52}\end{bmatrix} = N^{-1}G_u.$$

Define the sliding variables vector

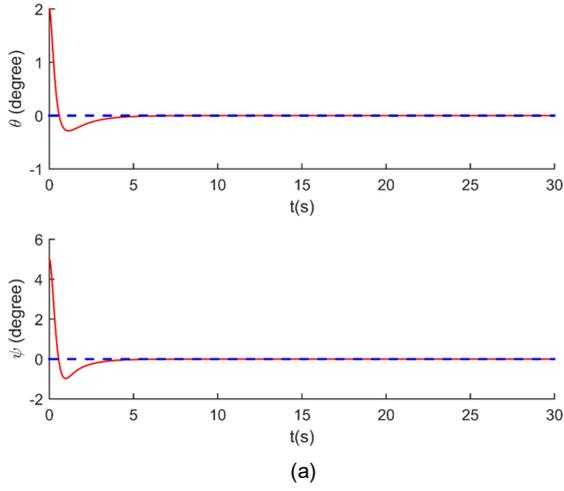

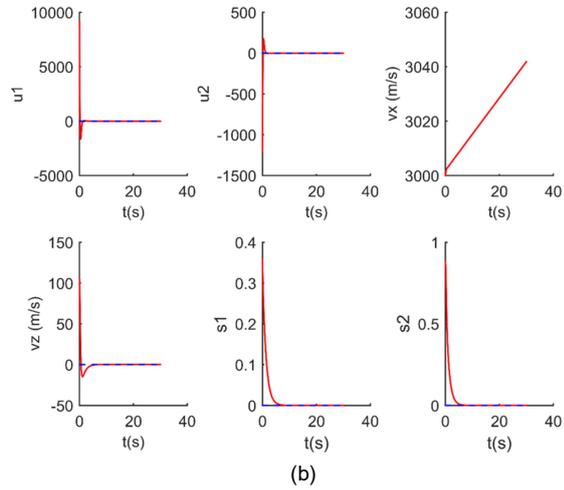

Fig. 2. First experiment using a general SMC as $k=1$. (a) Trajectories of states $\theta$ and $\psi$. (b) Control elements $u_1, u_2$; velocity $v_x$; state $v_z$; and sliding variables $s_1, s_2$.

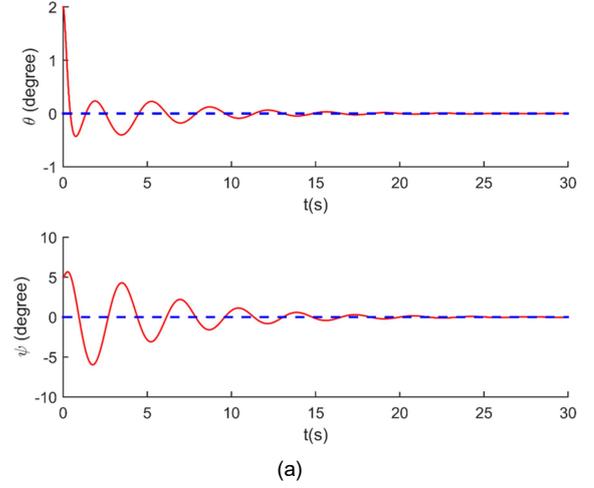

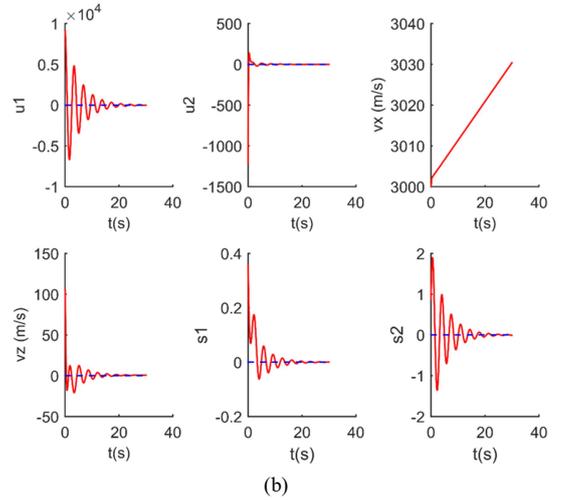

Fig. 3. First experiment using a general SMC as $k=7$. (a) Trajectories of states $\theta$ and $\psi$. (b) Control elements $u_1, u_2$; velocity $v_x$; state $v_z$; and sliding variables $s_1, s_2$.

$$s = \dot{e} + \lambda_1 e + \lambda_2 \int e(\tau) d\tau , \qquad (31)$$

where $s \in \mathbb{R}^2$ and $e = [\theta - \theta_d \quad \psi - \psi_d]^T$, $\theta_d$ is the target value of the attitude angle $\theta$, and $\psi_d$ is the target value of the fuel sloshing angle $\psi$; as mentioned above, $\theta_d = \psi_d = 0$, and $\lambda_1$ and $\lambda_2$ are strictly positive constants.

Differentiating (31), we get

$$\dot{s} = f(x) + G(x)u , \qquad (32)$$

where

$$f(x) = \begin{bmatrix} [A(x)]_3 \\ [A(x)]_5 \end{bmatrix} + \lambda_1 \begin{bmatrix} [x]_3 \\ [x]_5 \end{bmatrix} + \lambda_2 \begin{bmatrix} [x]_2 \\ [x]_4 \end{bmatrix},$$

$$G(x) = \begin{bmatrix} [B(x)]_{31} & [B(x)]_{32} \\ [B(x)]_{51} & [B(x)]_{52} \end{bmatrix}.$$

The system parametric uncertainties during flight could be formulated in the forms [4]

$$\begin{aligned} m^* &= m(1 + k\Delta m) \\ m_f^* &= m_f(1 + k\Delta m_f) \\ I_f^* &= I_f(1 + k\Delta I_f) \\ a^* &= a(1 + k\Delta a) \\ \varepsilon^* &= \varepsilon(1 + k\Delta \varepsilon) \end{aligned} \qquad (33)$$

where $m, m_f, I_f, a, \varepsilon$ are nominal values of parameters; $m^*, m_f^*, I_f^*, a^*, \varepsilon^*$ are real values; $\Delta m, \Delta m_f, \Delta I_f, \Delta a, \Delta \varepsilon$ express the basic uncertainties; and $k$ is a constant greater than or equal to 1. The basic uncertainty of each parameter satisfies the following boundedness hypothesis:

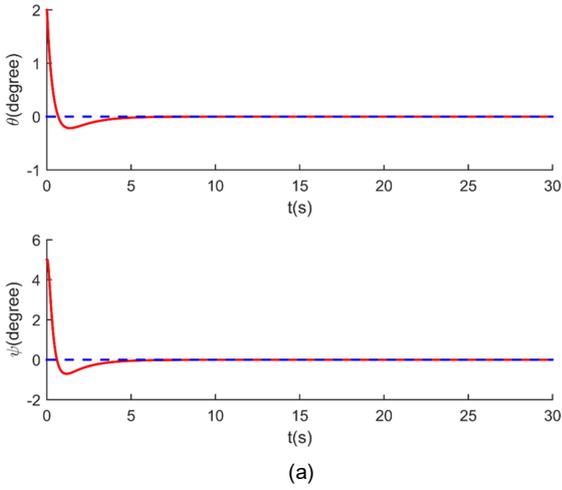

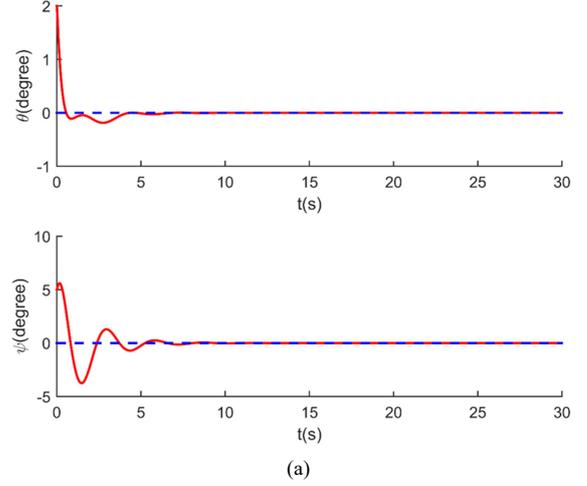

Fig. 4. Second experiment using our method as $k=1$. (a) Trajectories of states $\theta$ and $\psi$. (b) Control elements $u_1, u_2$, velocity $v_x$, state $v_z$, and sliding variables $s_1, s_2$.

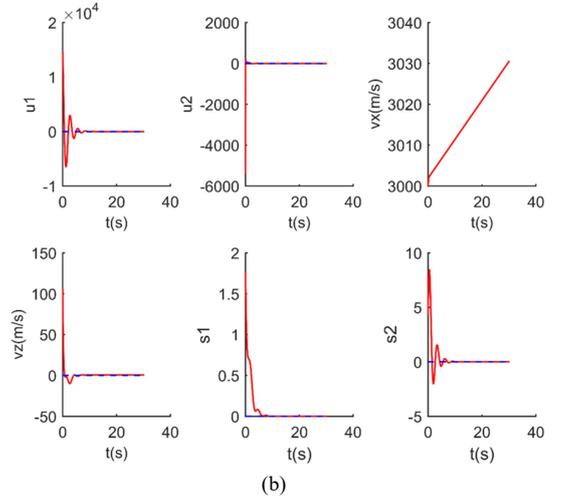

Fig. 5. Second experiment using our method as $k=7$. (a) Trajectories of states $\theta$ and $\psi$. (b) Control elements $u_1, u_2$, velocity $v_x$, state $v_z$, and sliding variables $s_1, s_2$.

$$\begin{aligned}|\Delta m| &\leq \overline{\Delta m} \\ |\Delta m_f| &\leq \overline{\Delta m_f} \\ |\Delta I_f| &\leq \overline{\Delta I_f} \\ |\Delta a| &\leq \overline{\Delta a} \\ |\Delta \varepsilon| &\leq \overline{\Delta \varepsilon}\end{aligned} \quad . \quad (34)$$

The upper bounds in (34) are known constants.

Due to the parametric uncertainties, $f(x)$ and $G(x)$ in (32) are not available. However, each of them can be separated into a known part and an unknown part. The unknown part of $f(x)$ satisfies Assumption 2, and its upper bound is $\bar{f}$. The known part of $G(x)$ is denoted by $G_0(x)$ and the unknown part by $\Delta G(x)$. By analysis, when the system state $x$ alters in a region containing the origin, and the parameter uncertainties are taken into consideration, the variation of $G_0(x)$ remains small; thus, referring to Remark 1, $G_0(x)$ can be chosen as a constant matrix denoted by $G_0$, and is calculated as the median value matrix of all possible $G(x)$. Then all the variations of $G(x)$ appear in $\Delta G(x)$.

Referring to Eqs. (8) and (9) and Remark 1, the PSO algorithm is applied to search a pair of matrices $M$, $Q$ such that $G_0 = MQ$ and the norm of the UBM of $F(x) = M^{-1}\Delta G(x)Q^{-1}$ (that is, $\|\overline{F}\|$ is minimized). According to the calculation process presented in Section 3.2, the control law is acquired.

*2) Simulation*

The comparison simulations are described to illustrate the performance of the proposed SMC and a general SMC. The objective of the first experiment was to test the robustness of the general SMC under different parametric uncertainties ($k=1$ and $k=7$). Referring to [4], the nominal values and upper bounds of system uncertain parameters are

$m = 600kg, m_f = 1000kg, I = 720 kg/m^2, I_f = 90 kg/m^2,$

$a = 0.32m, b = 0.25m, F = 1000N, \varepsilon = 0.0019, \overline{\Delta m} = 0.1,$
$\overline{\Delta m_f} = 0.1, \overline{\Delta I_f} = 0.05, \overline{\Delta a} = 0.05, \overline{\Delta \varepsilon} = 0.03$,
and the initial conditions of the spacecraft are
$v_{z0} = 105 m/s, \theta_0 = 2°, \dot{\theta}_0 = 0.57°/s, \psi_0 = 5°, \dot{\psi}_0 = 0.5°/s$, and
$v_{x0} = 3000 m/s$.

For the general SMC from [4], the control parameters remain unchanged for the two uncertain cases. The trajectories of $\theta$ and $\psi$ in different cases $k=1$ and $k=7$ are shown in Figs. 2(a) and 3(a) separately. It can be seen that as the degree of parametric uncertainty increases, the amplitude increases and the convergence time gets longer. Other variables described in Figs. 2(b) and 3(b) show the same trend as $k$ increases.

The second experiment was done to test the robustness of our method to different uncertainties. The system parameters, upper bounds, and initial conditions were as in the first experiment. The control parameters for both cases are listed below:

$\lambda_1 = 50, \lambda_2 = 125, \rho = 0.5, \overline{f} = [0.5 \quad 0.5]^T$,

$M = \begin{bmatrix} 0.17 & 2.97 \\ -4.63 & -2.32 \end{bmatrix} \times 10^{-2}, Q = \begin{bmatrix} 3.04 & 0.38 \\ 0.11 & 4.63 \end{bmatrix} \times 10^{-2}$,

$\overline{F} = \begin{bmatrix} 0.96 & 0.13 \\ 0.09 & 0.01 \end{bmatrix} (\|\overline{F}\|_2 = 0.97 < 1)$.

Figs. 4 and 5 show the system behaviors for different uncertain cases $k = 1$ and $k = 7$ respectively.

Figs 2(a) and 4(a) show that, in the case of a low degree of parametric uncertainty, two methods were able to stabilize the states $\theta$ and $\psi$ rapidly. The convergence times had little difference except that the overshoots of $\theta$, $\psi$, and $v_z$ in Figs. 4(a) and 4(b) are slightly smaller than those in Figs. 2(a) and 2(b) separately.

However, with a high degree of parametric uncertainty, the control performance shown in Fig. 5 is much better than that in Fig. 3. The state trajectories of $\theta$, $\psi$, and $v_z$ in Figs. 3(a) and 3(b) converge to zero with strong oscillatory behavior, whereas those in Figs. 5(a) and 5(b) converge in much less time and with much smaller amplitude.

*B. A two-link robot manipulator*
*1) Model analysis and controller design*
The dynamic equation of the two-link robot manipulator [21] shown in Fig. 6 is given by

$$M(q)\ddot{q} + C(q,\dot{q})\dot{q} + G(q) = \tau, \quad (35)$$

where $M(q)$ is the moment matrix of inertia, $C(q,\dot{q})\dot{q}$ the Coriolis centripetal torque vector, and $G(q)$ the gravitational torque vector. $q = [q_1 \quad q_2]^T$ and $q_1$, $q_2$ are the coordinates representing the angular positions. $\tau \in \mathbb{R}^2$ is the vector of applied torques. $q_d = [q_{1d} \quad q_{2d}]^T$, $q_{1d}$ and $q_{2d}$ are the target angular positions. For convenience, define $s_i = \sin(q_i)$ and $c_i = \cos(q_i)$, for $i = 1, 2$. The control objective is to control the robot angular positions to track the given target trajectories.

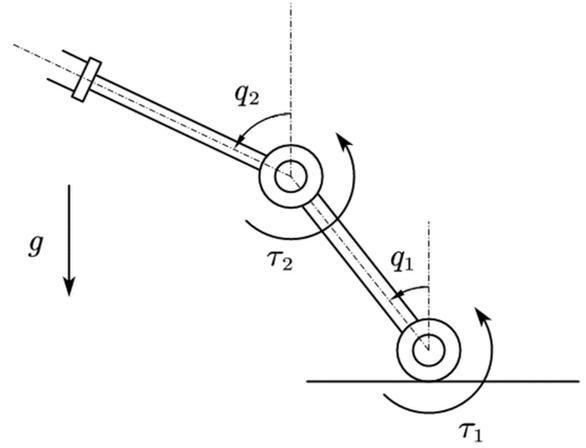

Fig. 6. Two-link robot manipulator dynamics

Define the vectors of state, target, and input as

$$x = [q_1 \quad \dot{q}_1 \quad q_2 \quad \dot{q}_2]^T, \quad (36)$$

$$x_d = [q_{1d} \quad \dot{q}_{1d} \quad q_{2d} \quad \dot{q}_{2d}]^T, \quad (37)$$

and

$$u = \tau. \quad (38)$$

Then the dynamic equation of the two-link manipulator in state-space form can be written as

$$\dot{x} = A(x) + B(x)u, \quad (39)$$

where

$[A(x)]_1 = [x]_2, [A(x)]_3 = [x]_4$,

$[A(x)]_2 = \dfrac{(s_1c_3 - c_1s_3) \times \left[ m_2 l_1 l_2 (s_1s_3 + c_1c_3)[x]_2^2 - m_2 l_2^2 [x]_4^2 \right]}{l_1 l_2 \left[ (m_1 + m_2) - m_2(s_1s_3 + c_1c_3)^2 \right]}$

$+ \dfrac{\left[ (m_1 + m_2)l_2 g s_1 - m_2 l_2 g s_3 (s_1s_3 + c_1c_3) \right]}{l_1 l_2 \left[ (m_1 + m_2) - m_2(s_1s_3 + c_1c_3)^2 \right]}$,

$[A(x)]_4 = \dfrac{(s_1c_3 - c_1s_3) \times \left[ -(m_1 + m_2)l_1[x]_2^2 + m_2 l_1 l_2 (s_1s_3 + c_1c_3)[x]_4^2 \right]}{l_1 l_2 \left[ (m_1 + m_2) - m_2(s_1s_3 + c_1c_3)^2 \right]}$

$+ \dfrac{\left[ -(m_1 + m_2)l_1 g s_1 (s_1s_3 + c_1c_3) + (m_1 + m_2)l_1 g s_3 \right]}{l_1 l_2 \left[ (m_1 + m_2) - m_2(s_1s_3 + c_1c_3)^2 \right]}$,

$[B(x)]_{11} = [B(x)]_{12} = [B(x)]_{31} = [B(x)]_{32} = 0$,

$[B(x)]_{21} = \dfrac{m_2 l_2^2}{m_2 l_1^2 l_2^2 \left[ (m_1 + m_2) - m_2(s_1s_3 + c_1c_3)^2 \right]}$,

$[B(x)]_{22} = \dfrac{-m_2 l_1 l_2 (s_1s_3 + c_1c_3)}{m_2 l_1^2 l_2^2 \left[ (m_1 + m_2) - m_2(s_1s_3 + c_1c_3)^2 \right]}$,

$[B(x)]_{41} = [B(x)]_{22}$,

$[B(x)]_{42} = \dfrac{(m_1 + m_2)l_1^2}{m_2 l_1^2 l_2^2 \left[ (m_1 + m_2) - m_2(s_1s_3 + c_1c_3)^2 \right]}$,

$m_1$ and $m_2$ are masses of the links, and $l_1$ and $l_2$ are lengths of the links.

Define the sliding variables vector as

$$s = \dot{e} + \alpha e, \quad (40)$$

where $s \in \mathbb{R}^2$, $e = \begin{bmatrix} [x]_1 - [x_d]_1 & [x]_3 - [x_d]_3 \end{bmatrix}^T$ is the tracking error vector, and $\alpha$ is a constant.

Differentiating (40), we get

$$\dot{s} = f(x) + G(x)u, \quad (41)$$

where

$$f(x) = \begin{bmatrix} [A(x)]_2 - \ddot{q}_{1d} \\ [A(x)]_4 - \ddot{q}_{2d} \end{bmatrix} + \alpha \begin{bmatrix} \dot{q}_1 - \dot{q}_{1d} \\ \dot{q}_2 - \dot{q}_{2d} \end{bmatrix},$$

$$G(x) = \begin{bmatrix} [B(x)]_{21} & [B(x)]_{22} \\ [B(x)]_{41} & [B(x)]_{42} \end{bmatrix}.$$

In this model, the uncertain parameters are addressing $m_1$, $m_2$, $l_1$, and $l_2$, and the ranges of uncertainty are bounded. Hence, $f(x), G(x)$ are not available. Referring to Assumption 2, the uncertain part of $f(x)$ is bounded by $\bar{f}$. $G_0$, the known part of $G(x)$, is calculated in a way similar to that of the first application.

Likewise, the PSO algorithm is used to search for a proper pair of matrices $M$ and $Q$ that satisfy $G_0 = MQ$ and makes $\|\bar{F}\|$ as small as possible. The control law is obtained on the basis of the calculation process presented in Section 3.2.

*2) Simulation*

The simulation results are included to illustrate the performance of the proposed SMC with much looser restrictions on the parameter uncertainties. As mentioned before, there are four uncertain parameters, and their values are limited within these ranges [21]:
$m_1 \in [0.7, 1.1]$, $m_2 \in [0.8, 1.4]$, $l_1 \in [0.9, 1.3]$, $l_2 \in [0.8, 1.3]$.

As in [21], the initial states are $x_0 = \begin{bmatrix} 0 & 0 & 0 & 0 \end{bmatrix}^T$, and the target state trajectories are selected as $q_{1d} = 0.01\sin(5t + \pi/2), q_{2d} = 0.01\sin(5t + \pi/2)$. The proposed controller is designed so that Assumption 3 is satisfied for arbitrarily evaluating the uncertain parameters within the given ranges. The control parameters are

$$M = \begin{bmatrix} -0.43 & -1.56 \\ 1.74 & 0.85 \end{bmatrix}, Q = \begin{bmatrix} -0.42 & 1.74 \\ -1.03 & 0.74 \end{bmatrix}, \bar{f} = \begin{bmatrix} 2 & 2 \end{bmatrix}^T,$$

$$\bar{F} = \begin{bmatrix} 0.79 & 0.20 \\ 0.18 & 0.68 \end{bmatrix} (\|\bar{F}\|_2 = 0.93 < 1).$$

Figs. 7 and 8 show the performance of our method in the following cases, which were adopted in [21]:

Case 1: $m_1 = 0.7kg, m_2 = 0.8kg, l_1 = 1.3m, l_2 = 1.3m$

Case 2: $m_1 = 1.1kg, m_2 = 1.4kg, l_1 = 0.9m, l_2 = 0.8m$.

As in [21], the angular positions $q_1$ and $q_2$ can rapidly track the target trajectories.

Simulations of our method in two other cases were carried out, and the results, presented in Figs. 9 and 10, also demonstrated satisfying performance. These cases were:

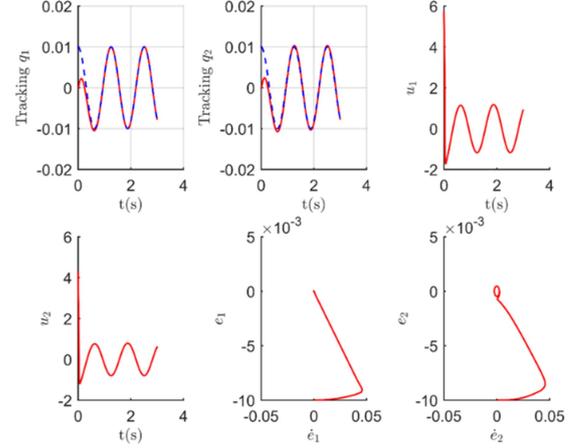

Fig. 7. Simulation results in the condition of Case 1

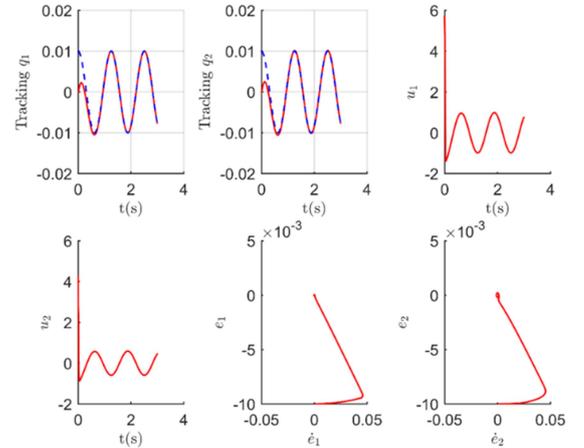

Fig. 8. Simulation results in the condition of Case 2

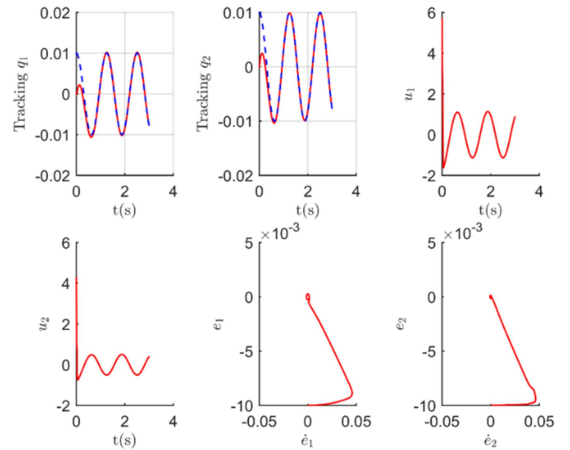

Fig. 9. Simulation results in the condition of Case 3

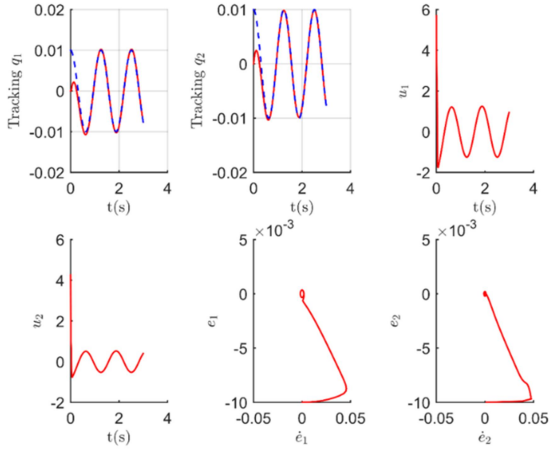

Fig. 10. Simulation results in the condition of Case 4

Case 3: $m_1 = 0.78kg, m_2 = 0.96kg, l_1 = 1.18m, l_2 = 0.81m$
Case 4: $m_1 = 0.76kg, m_2 = 0.91kg, l_1 = 1.21m, l_2 = 0.85m$.

However, the condition in [21] was not met in both cases, because the UBM of matrix $\Delta G(x)G_0^{-1}(x)$ are $\begin{bmatrix} 0.54 & 0.39 \\ 0.79 & 1.02 \end{bmatrix}$ and $\begin{bmatrix} 0.65 & 0.40 \\ 0.91 & 1.06 \end{bmatrix}$ respectively; some elements of them were much greater than $1/2$, then the condition in [21], which is mentioned earlier in Section 2, was unsatisfied.

## V. CONCLUSION

*This paper presents a new SMC algorithm that stabilizes a class of affine dynamical systems, with the corresponding HFGM being neither deterministic nor positive definite. The proposed approach allows one to obtain, with a low computational burden, a control law by solving a nonlinear vector equation instead of the conventional algebraic expression, and the existence and uniqueness of the solution are theoretically guaranteed. The chattering caused by the discontinuous control law can be suppressed easily, as it can in general SMCs. The stability of the closed-loop system is formally analyzed, and the validity of the approach is investigated using two applications addressing the controls of a liquid-filled spacecraft and a two-link robot manipulator.*

## APPENDIX A

**Proof of Lemma 1**
**Proof.** Assume that an $m$-dimensional vector $y$ belongs to both $C_i^o$ and $C_j^o$, $i \neq j$; that is, there exists $p_i, p_j > 0$ such that $y = D_i p_i = D_j p_j$. It follows that

$$S_j p_j - S_i p_i = H(p_i - p_j) \quad (42)$$

If $p_i = p_j$, (42) is deduced to $S_j = S_i$ which means that $C_j^o$ and $C_i^o$ are the same cone.

If $p_i \neq p_j$, by using the condition $\|H\| < 1$, the norm of the left side of (42) satisfies the inequality $\|S_j p_j - S_i p_i\| \leq \|H\| \|p_i - p_j\| < \|p_i - p_j\|$. However, because $p_1, p_2$ are positive vectors, the vector inequality $|S_i p_i - S_j p_j| \geq |p_i - p_j|$ holds, thus the norm of the left side of (42) satisfies $\|S_i p_i - S_j p_j\| \geq \|p_i - p_j\|$. This contradiction indicates that $y$ cannot belong to different open convex cones $C_i^o$ and $C_j^o$ simultaneously. Lemma 1 is proved. □

## APPENDIX B

**Proof of Lemma 2**
**Proof.** Let $y \in \hat{C}_i$. According to Definition 1, $y = (S_i + H)p$ is with at least one element of $p$ being zero. Then there exists $S_j \neq S_i$ so that $S_j p = S_i p$. Therefore, $y = (S_j + H)p$ holds too, so $y$ also belongs to $\hat{C}_j$; consequently, the conical surface $\hat{C}_i$ is not independent. Lemma 2 is proved. □

## APPENDIX C

**Proof of Lemma 3**
**Proof.** $C \subset \mathbb{R}^m$ is the union of finite closed convex cones, so $C$ is a closed cone. According to Lemma 1, the conical surface of $C$ is the closure of union of the independent conical surface of $C_i, i \in [2^m]$. But according to Lemma 2, there is no independent conical surface for any $C_i$, so the conical surface of $C$ is an empty set. However, $\mathbb{R}^m$ itself is a closed cone with no conical surface, so the result can only be $C = \mathbb{R}^m$. □

ACKNOWLEDGMENT

The research is supported by the National Natural Science Foundation of China under Grant no. 61673312.

**Zu-Ren Feng** (M'05) received M. Eng. and Ph.D. degrees in information and control engineering from Xi'an Jiaotong University, Xi'an, China, in 1982 and 1988, respectively. Since 1994, he has been a Professor with the School of Electronic and Information Engineering, Xi'an Jiaotong University. In 1992, he worked as a visiting scholar in INRIA, France, for research on manipulator control with flexible joints and applications of Petri Nets in DEDS. In 1994, he was invited by Kassel University, Germany, for research on mobile service robots. In 2006 and 2007, he worked as a visiting professor in the University of Bradford, U.K. for research on multi-agent systems. His research interests include robotics and automation, intelligent information processing, and evolutionary computing based optimization.

**Rui-Zhi Sha** received her B.S.degree in Automation from Xi'an Jiaotong University, Xi'an, China in 2014. She is currently pursuing the Ph.D. degree in control science and engineering from the School of Electronic and Information Engineering, Xi'an Jiaotong University, Xi'an, China. Her research interests include nonlinear systems control, robotics automation.

**Na Lu** received her B.S. and Ph.D. degrees from Xi'an Jiaotong University, Xi'an, Shaanxi, China in 2002 and 2008, respectively. Currently, she is an Associate Professor at Xi'an Jiaotong University, Xi'an, Shaanxi, China. Her research interests include statistical image analysis, machine learning, cognitive science and robotics.

**Chen-Long He** received his B.S.degree in vehicle and information engineering, in 2006 and M.S. degrees in vehicle application engineering, in 2009 from Southwest Jiaotong

University, Chengdu, China. He is currently pursuing the Ph.D. degree in control science and engineering from the School of Electronic and Information Engineering, Xi'an Jiaotong University, Xi'an, China. His research interests include mobile sensor networks, multi-robot systems, and collective motion of self-propelled particles.